\documentclass[pdflatex,sn-mathphys-num,referee]{sn-jnl}

\usepackage{graphicx}%
\usepackage{multirow}%
\usepackage{amsmath,amssymb,amsfonts}%
\usepackage{amsthm}%
\usepackage{mathrsfs}%
\usepackage[title]{appendix}%
\usepackage{xcolor}%
\usepackage{textcomp}%
\usepackage{manyfoot}%
\usepackage{booktabs}%
\usepackage{algorithm}%
\usepackage{algorithmicx}%
\usepackage{algpseudocode}%
\usepackage{listings}%
\usepackage[latin1]{inputenc}%
\usepackage{subcaption}

\theoremstyle{thmstyleone}%
\theoremstyle{thmstyletwo}%

\theoremstyle{thmstylethree}%

\begin{document}

\title[Quantum Machine Learning Applied to the Sinking of the Titanic]{Quantum Machine Learning Applied to the Sinking of the Titanic}


\author[1]{\fnm{Luiz Henrique} \sur{Prudencio dos Santos}}

\author[1]{\fnm{Eliane F.} \sur{Chinaglia}}

\author[1]{\fnm{Jessica} \sur{Fleury Curado}}

\author[1]{\fnm{Marcilei A.} \sur{Guazzelli}}

\author[1]{\fnm{Mariana} \sur{Pojar}}

\author[1]{\fnm{Sueli} \sur{Hatsumi Masunaga}}

\author*[1]{\fnm{Roberto} \sur{Baginski Batista Santos}}\email{rsantos@fei.edu.br}


\affil[1]{\orgdiv{Physics Department}, \orgname{Centro Universit\'{a}rio FEI}, \orgaddress{\street{Av. Humberto A. C. Branco, 3972}, \city{S\~{a}o Bernardo do Campo}, \postcode{09850-901}, \state{SP}, \country{Brazil}}}


\abstract{This work investigates the performance of hybrid quantum-classical variational classifiers applied to a supervised learning task involving the \texttt{titanic3} dataset. Quantum models were constructed using Pauli entangling and non-entangling expansion-based feature maps and the \texttt{RealAmplitudes} ansatz with up to $50$ variational parameters. Model training employed the COBYLA gradient-free optimizer to minimize the cross-entropy loss, within an ideal statevector simulation framework. Comparative performance analysis reveals that the models based on the non-entangling feature map consistently outperformed the models based on the entangling features maps, achieving saturation of classification metrics (accuracy, balanced accuracy, and Youden's index) beyond $15$ to $20$ parameters. Further, two quantum models were benchmarked against a classical Support Vector Classifier (SVC). While both approaches yielded similar predictive performance across multiple training sizes, the classical model exhibited a performance collapse when trained with $90\%$ of the dataset, a failure mode absent in the quantum classifiers. These results underscore the robustness and viability of variational quantum classifiers for binary classification tasks on classical datasets in the NISQ era.}




\maketitle

\section{Introduction}\label{sec1}

Quantum computing aims to explore some features unique to Quantum Mechanics to solve computational problems. \textit{Circa} 1970, Wiesner devised the conjugate coding, a class of quantum communication protocols as an example of the restrictions imposed by the uncertainty principle on measurement \cite{Wiesner:1983}. Counterfeit-proof quantum money and a protocol for sending two mutually exclusive messages are examples of conjugate coding. Building on these pioneering ideas, Bennett and Brassard presented the so-called BB84 protocol for quantum key distribution \cite{Bennett+Brassard:1984}. A few years later, Ekert introduced an entanglement-based quantum key distribution protocol \cite{Ekert:1991}.

The no-cloning theorem was discovered \cite{Park:1970}, and rediscovered a dozen years later \cite{Wootters+Zurek:1982,Dieks:1982,Milonni+Hardies:1982}. Feynman discussed the need for a quantum computer in order to efficiently simulate physical systems \cite{Feynman:1982,Feynman:1986}, Benioff developed the concept of a Quantum Turing Machine \cite{Benioff:1980,Benioff:1982a,Benioff:1982b,Benioff:1982c}, laying the foundations upon which Deutsch built the concept of a Universal Quantum Computer \cite{Deutsch:1985}.

Soon after that, quantum algorithms as Deutsch's \cite{Deutsch:1985}, Deutsch-Josza \cite{Deutsch+Jozsa:1992}, superdense coding \cite{Bennett+Wiesner:1992}, Bernstein-Vazirani \cite{Bernstein+Vazirani:1993,Bernstein+Vazirani:1997}, quantum teleportation \cite{Bennett+etal:1993}, Simon's algorithm \cite{Simon:1994,Simon:1997}, Shor's discrete logarithm and prime factorization algorithms \cite{Shor:1994,Shor:1997}, the Quantum Fourier Transform \cite{Coppersmith:1994}, quantum phase estimation \cite{Kitaev:1995}, Grover's search algorithm \cite{Grover:1996,Grover:1997}, quantum amplitude amplification and estimation \cite{BHMT:2002}, and the HHL algorithm to solve linear systems \cite{Harrow+Hassidim+Lloyd:2008} indicated that quantum computing could present an advantage over classical computing regarding specific problems.

Quantum machine learning (QML) is the branch of quantum computing that attempts to harness the unique characteristics of quantum mechanics to improve machine learning (ML) tasks \cite{Aimeur+Brassard+Gambs:2006,Lloyd+Mohseni+Rebentrost:2013,Wittek:2014,Schuld+Sinayskiy+Petruccione:2015,Biamonte+etal:2017,Dunjko+Briegel:2018,Benedetti+etal:2019,Havlicek+etal:2019,Schuld+Petruccione:2021,Cerezo+etal:2021,Cerezo+etal:2022}. Typical machine learning tasks are dimensionality reduction, classification, regression, clustering, and density estimation , all of which tries to make predictions generalizing from patterns found in data \cite{Deisenroth+Faisal+Ong:2020,Schuld+Petruccione:2021}. Machine learning is one of the most rapidly growing fields of the last few decades \cite{Sevilla+etal:2022}, and there are expectations on the speed-ups that machine learning could get from exploring quantum resources such as superposition, entanglement, coherence, or nonlocality \cite{Dunjko+Briegel:2018}. We will follow the narrow definition of quantum machine learning as ``machine learning with quantum computers" \cite{Schuld+Petruccione:2021}.

In this paper, we will restrict ourselves to the classification problem in supervised machine learning, which may be defined in the following way for classical data. A training data set $T=\{(\vec{x}_0,y_0),\cdots,(\vec{x}_{N-1},y_{N-1})\}$ composed of $N$ instances $\vec{x}_i\in\mathbb{R}^D$ to which are associated the labels $y_i\in \{0,1\}$ is analyzed to predict the label $\hat{y}'$ associated with a new instance $\vec{x}'\in \mathbb{R}^D$, that is, to classify an instance that was not part of the data set previously analyzed during the training phase of the classifier. It is assumed that there is a correlation between the ideal classification and the classification inferred from the training set. The dimensionality $D$ of an instance $\vec{x}=(x_0,x_1,\cdots,x_{D-1})$ corresponds to the number of features present in the data. In practice, a larger dataset $\Omega\subset\mathbb{R}^D$ is splitted in a training set $T$ and a test set $T'$, and the test set instances are used to assess the classifier prediction quality before the classifier model is deployed.

A quantum machine learning task with classical data begins with the encoding of the classical data into the registers of a quantum computer. Loading classical data into a quantum computer alone takes linear time, ruling out the possibility of more aggressive speedups \cite{Schuld+Petruccione:2021}. There is not consensus on how to encode data into quantum states. Quite generally, an encoding falls into the class of feature maps, in which a nonlinear transform is applied to the data leading to a quantum state $|\Phi(\vec{x})\rangle=\mathcal{U}_{\Phi(\vec{x})}|0\rangle$ \cite{Havlicek+etal:2019,Schuld+Petruccione:2021}. Less general feature maps such as basis encoding, amplitude encoding, angle encoding, and Pauli expansion circuits are commonly used in the field \cite{Havlicek+etal:2019,Schuld+Petruccione:2021}. Considerations on the number of qubits, expressiveness of the encoding, circuit depth, noise level of the target quantum device, and the runtime may influence the choice of a specific encoding for a given dataset.

Besides encoding data, the model needs to learn some pattern about the data. A variational quantum circuit (VQC), a circuit containing quantum gates that depend on a set $\vec{\theta}=\{\theta_0, \theta_1,\cdots,\theta_k\}$ of $k$ parameters, may be used to transform the state $|\Phi(\vec{x})\rangle$ into the state $|\psi_{\vec{\theta}}(\vec{x})\rangle = \mathcal{U}_{\vec{\theta}}|\Phi(\vec{x})\rangle$. Considerations on the number of qubits, gates, connectivity and noise level of the target quantum device, circuit depth, the complexity and dimensionality of the data, the balance between expressivity and trainability, and domain knowledge may guide the choice of the VQC for a particular task. Following the time-honored tradition of the variational method in Quantum Mechanics, the VQC is frequently called an ansatz.

In this model, learning occurs while the parameters $\vec{\theta}$ are varied in order to minimize a loss function $\ell(\vec{\theta})$ with respect to the parameters $\vec{\theta}$. For present-day noisy intermediate-scale quantum computers (NISQ) \cite{Preskill:2018}, this optimization is performed classically.

Optimization algorithms may be local or global, gradient-based or gradient-free. Local optimization is computationally efficient, but the solution may be just a sub-optimal local minimum in the parameter landscape, instead of a global minimum. Moreover, the solution found by local optimization may be highly dependent on the initial position of the system in the parameter landscape. In contrast, global optimization tries to ensure that a global minimum is found, but it may be computationally expensive.

A gradient-based optimization algorithm uses the gradient of the loss function to iteratively search for the minimum of that function. Although it is possible to estimate or even to analytically determine the gradient of a variational quantum circuit \cite{Schuld+etal:2019}, it may be difficult. Gradient-free optimization algorithms avoid the need for this costly gradient evaluation, relying on loss differences $\ell(\vec{\theta}_B)-\ell(\vec{\theta}_A)$ between points in the parameter landscape. However, the barren plateau problem, in which the parameter landscape becomes flat as the problem size increases affects both gradient-based as well gradient-free optimization algorithms, turning the search into a random walk through a exponentially flat region in the parameter landscape, and reducing the trainability of the model \cite{Arrasmith+etal:2021,Larocca+etal:2025}. The origin of the barren plateaus is attributed to circuit expressiveness, quantum hardware noise, and misalignment between the encoded state $|\psi_{\vec{\theta}}(\vec{x})\rangle$, the VQC, and the measurement \cite{Larocca+etal:2025}. Limiting the expressiveness of the circuit by reducing its depth and number of qubits may mitigate the problem in quantum machine learning, but the landscape of shallow models may have many local minima very above the global minimum, making training difficult \cite{Anschuetz+etal:2022}.

\begin{figure}[htp]
\centering
\includegraphics[width=\textwidth]{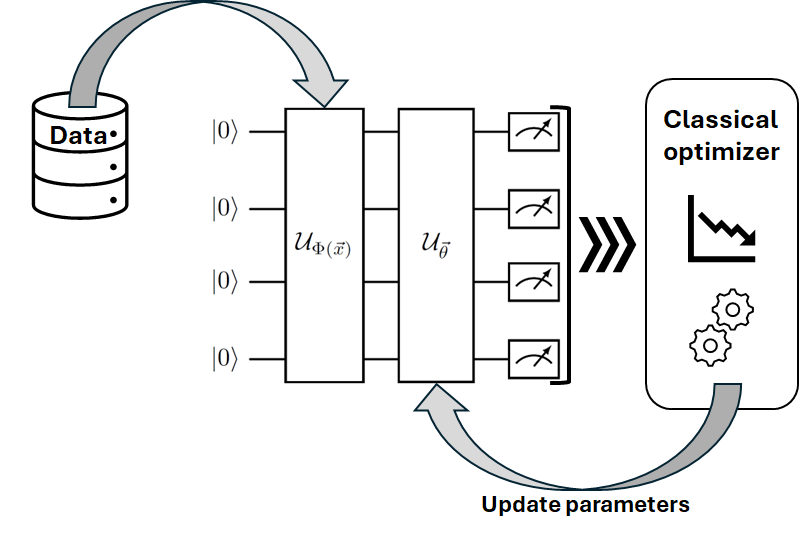}
\caption{Illustration of a hybrid quantum-classical machine learning variational model.}\label{fig:HQCMLVM}
\end{figure}

The model described above, illustrated in Figure~\ref{fig:HQCMLVM}, is a hybrid quantum-classical machine learning variational model, which seems to be useful in the context of today's limited noisy intermediate-scale quantum computers \cite{Benedetti+etal:2019}. Improvements in hardware quality, error mitigation and correction, and quantum software design may open the way to models that could be very different from this.

In this paper, we present an application of quantum machine learning to the classification of the well known \texttt{titanic3} dataset \cite{Harrell:2001,Harrell+Cason:2002}. This particular dataset contains data about 1309 passengers of the RMS Titanic ship, including their survival status, and additional 13 features such as name, age, sex, port of embarkation (Cherbourg, Queenstown, now Cobh, or Southampton), passenger class (1st, 2nd, or 3rd), number of siblings/spouses aboard, number of parents/children aboard, ticket number, fare, cabin number, lifeboat number (if the passenger survived), body number (if the passenger perished, and their body was recovered), and home destination. Approximately $6,5\%$ of the data is missing, especially regarding the features age and cabin number. This dataset is widely used as a paradigmatic example in supervised learning; in fact, a competition based on the dataset is ongoing at the Kaggle website since 2012 with almost 16 thousand entries \cite{kaggle:2012}.

\section{Methods}\label{sec2}

An exploratory data analysis was performed on the \texttt{titanic3} dataset. Passenger's name, ticket, and port of embarkation were deemed irrelevant with respect to their survival status, and dropped. On the other hand, a non-missing lifeboat number or a non-missing body number were equivalent to ``survived" and to ``perished", respectively, so they were also dropped. In the dataset, the fare reported may be the fare charged from a single individual or from an entire family; hence we opted to drop this feature owing to this methodological inconsistency. The features cabin and home destination were also dropped because they contained many missing values. Information on age was missing for 263 passengers, and, for these passengers, we imputed the mean age as their age, in an effort to preserve a potentially relevant feature.

After this data cleanup, we were left with five features of $1309$ passengers: ``pclass" (passenger class), ``sex", ``age", ``sibsp" (number of siblings or spouses aboard), and ``parch" (number of parents or children aboard). All features were scaled to the $[0,1]$ range by the class \texttt{MinMaxScaler} of the Python library scikit-learn \cite{scikit-learn:2011}. The dataset is imbalanced since it contains $500$ passengers who survived the sinking and $809$ passengers who perished in the sinking.

We built several quantum classifiers composed of combinations of the Pauli expansion circuits \texttt{ZFeatureMap} and \texttt{ZZFeatureMap} as feature maps and the \texttt{RealAmplitudes} ansatz with different number of parameters. \texttt{ZFeatureMap} and \texttt{ZZFeatureMap} are sub-classes of the Pauli expansion circuit \cite{Havlicek+etal:2019}. In the first-order expansion \texttt{ZFeatureMap}, after a layer of Hadamard $\mathrm{H}$ gates that puts every qubit in the $|+\rangle = (|0\rangle + |1\rangle)/\sqrt{2}$ state, classical data is encoded in the quantum registers through a layer of phase gates $\mathrm{P}(\varphi)$, where the phase angle $\varphi$ is a linear function of the feature value being encoded. In the second-order expansion \texttt{ZZFeatureMap}, the phase angles depends linearly and non-linearly on the feature values. Besides that, there are no entangling gates in the first-order expansion \texttt{ZFeatureMap}, while the second-order expansion \texttt{ZZFeatureMap} presents several entanglement gates through the \texttt{Controlled NOT} ($\mathrm{cX}$) gate. Both \texttt{ZFeatureMap} and \texttt{ZZFeatureMap} are available as classes of the Python library Qiskit \cite{qiskit:2024}.

\begin{figure}[htp]
\centering
\includegraphics[width=\textwidth]{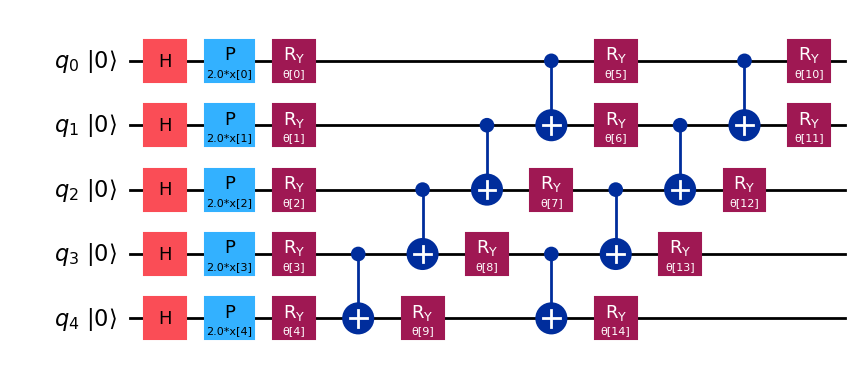}
\caption{A five-qubit classifier composed of the non-entangling \texttt{ZFeatureMap} feature map and the \texttt{RealAmplitudes} ansatz with 15 variational parameters.}\label{fig:classifier}
\end{figure}

The \texttt{RealAmplitudes} ansatz is composed by alternating layers of rotations $\mathrm{R_y}(\theta)$ and entanglement gates $\mathrm{cX}$. The rotation angles are the variational parameters that must be optimized in order for the model to learn. The number of repetitions of the basic cell defines the number of parameters of the model. This ansatz is also provided by the Qiskit library. Figure~\ref{fig:classifier} shows a five-qubit classifier with 15 variational parameters while Table~\ref{tab:1} presents the quantum classifiers whose performance was evaluated in this paper.

\begin{table}[h]
\caption{Models evaluated in this paper}\label{tab:1}%
\begin{tabular}{@{}rlc@{}}
\toprule
Classifier name & Feature Map & Number of variational parameters \\
\midrule
\texttt{Z10}    & \texttt{ZFeatureMap}   & $10$ \\
\texttt{Z15}    & \texttt{ZFeatureMap}   & $15$ \\
\texttt{Z20}    & \texttt{ZFeatureMap}   & $20$ \\
\texttt{Z25}    & \texttt{ZFeatureMap}   & $25$ \\
\texttt{Z30}    & \texttt{ZFeatureMap}   & $30$ \\
\texttt{Z35}    & \texttt{ZFeatureMap}   & $35$ \\
\texttt{Z40}    & \texttt{ZFeatureMap}   & $40$ \\
\texttt{Z45}    & \texttt{ZFeatureMap}   & $45$ \\
\texttt{Z50}    & \texttt{ZFeatureMap}   & $50$ \\
\texttt{ZZ10}   & \texttt{ZZFeatureMap}  & $10$ \\
\texttt{ZZ15}   & \texttt{ZZFeatureMap}  & $15$ \\
\texttt{ZZ20}   & \texttt{ZZFeatureMap}  & $20$ \\
\texttt{ZZ25}   & \texttt{ZZFeatureMap}  & $25$ \\
\texttt{ZZ30}   & \texttt{ZZFeatureMap}  & $30$ \\
\texttt{ZZ35}   & \texttt{ZZFeatureMap}  & $35$ \\
\texttt{ZZ40}   & \texttt{ZZFeatureMap}  & $40$ \\
\texttt{ZZ45}   & \texttt{ZZFeatureMap}  & $45$ \\
\texttt{ZZ50}   & \texttt{ZZFeatureMap}  & $50$ \\
\botrule
\end{tabular}
\end{table}

We used the COBYLA Optimizer (Constrained Optimization by Linear Approximation) \cite{Powell:1994}. COBYLA is a simplex gradient-free local optimization algorithm based on linear polynomial approximations to the loss function. Qiskit Machine Learning library \cite{qiskit-machine-learning:2025} provides a convenient interface to the COBYLA method of the \texttt{optimize} class of the Python library SciPy \cite{scipy:2020}. Among other optimization methods easily available through Qiskit Machine Learning library, we highlight ADAM, Gradient Descent, Nelder-Mead, and SPSA. Although some optimizers provide several options to control their behavior, we have chosen COBYLA since it is a relatively simple algorithm that works well with very little tuning.

In classification tasks in machine learning, a loss function represents a measure of the difference between two probability distributions, the ground truth probability distribution implied by the labels of the dataset and the probability distribution predicted by the model. We selected the cross-entropy given by

\begin{equation}
 H(p,q) = - \sum_c p_c \log(q_c)
 \label{eq:1}
\end{equation}

\noindent as a loss function, where $p_c$ is the probability of the class $c$ in the dataset, and $q_c$ is the probability of the class $c$ predicted by the model. The difference between the cross-entropy $H(p,q)$ and the entropy

\begin{equation}
 H(p) = - \sum_c p_c \log(p_c)
 \label{eq:2}
\end{equation}

\noindent is the Kullback-Leibler divergence, or the relative entropy of the distribution $p$ with respect to the distribution $q$, given by

\begin{equation}
 D_\mathrm{KL}(p\|q) = H(p,q) - H(p) = - \sum_c p_c \log\left(\frac{p_c}{q_c}\right)\geq 0.
 \label{eq:3}
\end{equation}

The cross-entropy is commonly used as a loss function in classification tasks in machine learning and the Qiskit Machine Learning library provides an interface to access this function from the SciPy library.

\begin{figure}[htp]
\centering
\includegraphics[width=\textwidth]{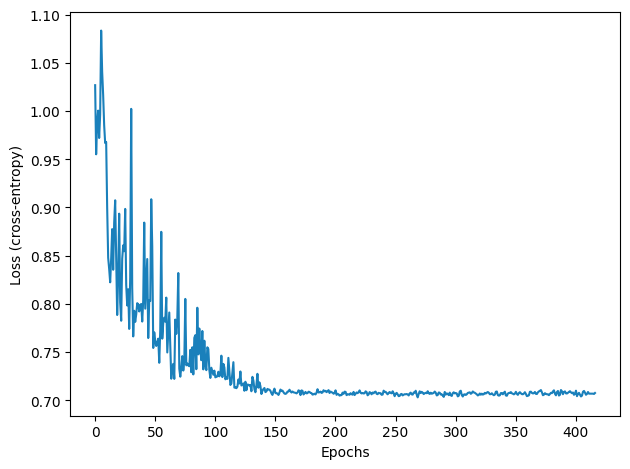}
\caption{Cross-entropy during the training of the \texttt{Z35} model with $70\%$ of the dataset.}\label{fig:cross-entropy}
\end{figure}

Initially, we trained the model \texttt{Z35} with $70\%$ of the data from the dataset for $418$ epochs. Figure~\ref{fig:cross-entropy} shows the cross-entropy during this training. After approximately $150$ epochs, the cross-entropy seems to converge. This result led us to set $150$ epochs as the standard for the training of all models in this paper.

All classifiers were trained with a fraction of $70\%$ of the data contained in the dataset with $150$ epochs. To compare their performance, we tested the predictions of each model with the remaining $30\%$ of the dataset. After the training and testing phases, the predictions of each model were compared to the ground truth, and the elements of the contingency table (confusion matrix) were tallied. In the binary classification case, these elements correspond to the number of true positive ($\mathrm{TP}$), true negative ($\mathrm{TP}$), false positive ($\mathrm{FP}$), and false negative ($\mathrm{FN}$) cases. A prediction is classified as true positive (negative) when the model correctly predicts that the passenger has survived (perished); it is classified as false positive (negative) when the model incorrectly predicts that the passenger has survived (perisehd). The number of positive cases in a given dataset is $\mathrm{P}=\mathrm{TP}+\mathrm{FN}$ while the number of negative cases is $\mathrm{N}=\mathrm{TN}+\mathrm{FP}$.

Performance metrics such as the positive predictive value $\mathrm{PPV}=\mathrm{TP}/(\mathrm{TP}+\mathrm{FP})$, also known as the precision of the positive class, the true positive rate $\mathrm{TPR}=\mathrm{TP}/\mathrm{P}$, also known as the sensitivity or recall of the positive class, the negative predictive value $\mathrm{NPV}=\mathrm{TN}/(\mathrm{TN}+\mathrm{FN})$, also known as the precision of the negative class, the true negative rate $\mathrm{TNR}=\mathrm{TN}/\mathrm{N}$, also known as the specificity or the recall of the negative class, the accuracy $\mathrm{ACC}=(\mathrm{TP}+\mathrm{TN})/(\mathrm{P}+\mathrm{N})$, the balanced accuracy $\mathrm{bACC}=(\mathrm{TPR}+\mathrm{TNR})/2$, and the Youden's index $J=\mathrm{TPR}-(1-\mathrm{TNR})$ were used to assess the predictive power of the models \cite{Sokolova+Lapalme:2009,Tharwat:2021}. Training and test were performed in a Google Compute Engine back-end (Google Colab) using the \texttt{Statevector} ideal simulator method from the Qiskit library.

Finally, two of the best performance quantum classifiers with few parameters were selected and compared with the classical support vector classifier SVC with linear kernel from the scikit-learn library. The linear SVC is a classifier based on a support vector machine (SVM), in which the model has to learn a set of parameters defining a hyperplane that separates the classes while maximizing the margin \cite{Cortes+Vapnik:1995,Brunton+Kutz:2022}. The margin is the distance between the support vectors, a set of vectors of each class which are closest to the separating hyperplane. Accuracy $\mathrm{ACC}$, balanced accuracy $\mathrm{bACC}$, and Youden's index $J$ were used as performance metrics.

\section{Results and discussion}\label{sec3}

\subsection{Comparison among quantum models}\label{subsec3.1}

\begin{figure}[htp]
\includegraphics[width=\textwidth]{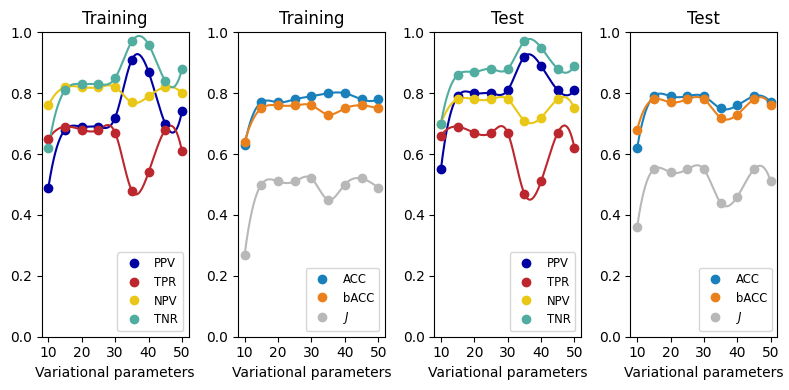}
\caption{Performance metrics of the \texttt{ZFeatureMap}-based models as a function of the number of variational parameters of the model.}\label{fig:Z_metrics}
\end{figure}

Figure~\ref{fig:Z_metrics} shows performance metrics for the \texttt{ZFeatureMap}-based classifiers as a function of their variational parameter number in the training and in the test phases. Analysis of the accuracy, balanced accuracy and Youden's index shows that the performance of these models saturates once the models reach $15$ or $20$ parameters. Accuracy and balanced accuracy saturates a bit below $0.80$, indicating that these models display a good ability to label instances of this dataset correctly even with minimal data cleaning, feature engineering or kernel engineering.

\begin{figure}[htp]
\includegraphics[width=\textwidth]{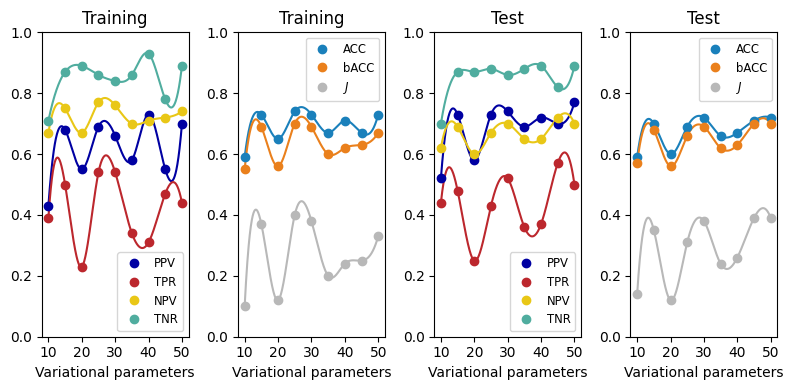}
\caption{Performance metrics of the \texttt{ZZFeatureMap}-based models as a function of the number of variational parameters of the model.}\label{fig:ZZ_metrics}
\end{figure}

Figure~\ref{fig:ZZ_metrics} shows performance metrics for the \texttt{ZZFeatureMap}-based classifiers as a function of their variational parameter number in the training and in the testing phases. Accuracy and balanced accuracy fluctuates roughly in the range $0.60--0.75$, and the low values of the Youden's indices indicate that \texttt{ZZFeatureMap}-based models faced difficulties classifying correctly and avoiding failure in the classification.

\begin{figure}[htp]
\centering
\includegraphics[width=0.7\textwidth]{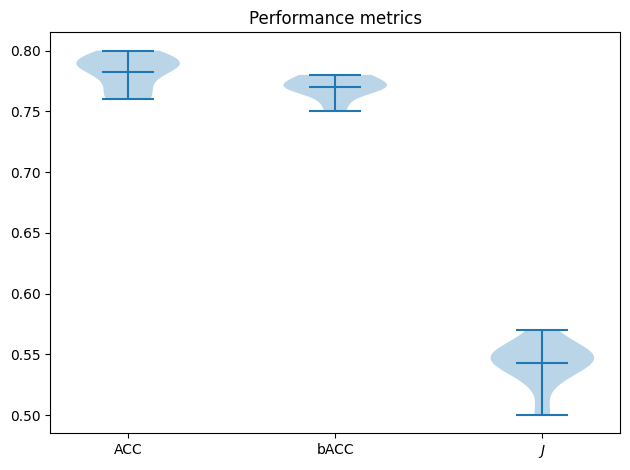}
\caption{Violin plots for accuracy $\mathrm{ACC}$, balanced accuracy $\mathrm{bACC}$, and Youden's index $J$ of $10$ instances of the \texttt{Z20} quantum model.}\label{fig:Violin_plots}
\end{figure}

Since the results obtained with a quantum model are not deterministic, we trained and tested $10$ different instances of the \texttt{Z20} model in order to estimate the fluctuations affecting their accuracy, balanced accuracy, and Youden's index. The instances were all trained with the same $70\%$ of the data and they were tested on the same remaining $30\%$ of the data. This way, the only non-deterministic factors were the initial values of the parameters, the optimizer search path, and the qubits measurements. Figure~\ref{fig:Violin_plots} present violin plots for the distribution of accuracy $\mathrm{ACC}$, balanced accuracy $\mathrm{bACC}$, and Youden's index $J$ obtained in these tests. Inspection of the plots indicates that these non-deterministic factor did not affected significantly the results.

\subsection{Comparison with a classical classifier}\label{subsec3.2}

\begin{figure}[htp]
\includegraphics[width=\textwidth]{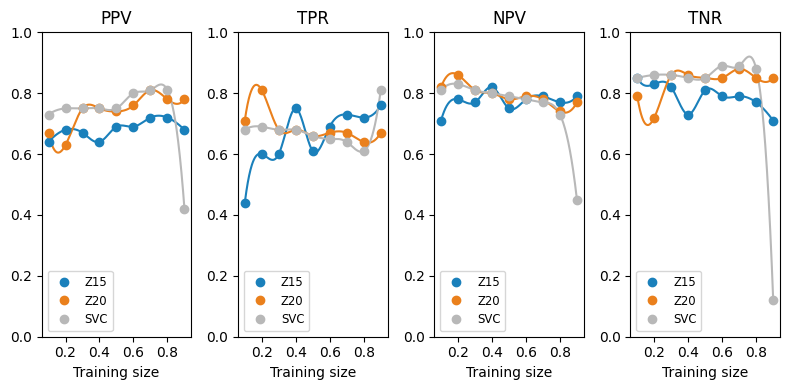}
\caption{Comparison of the performance of the \texttt{Z15}, \texttt{Z20} quantum models, and the \texttt{SVC} classical model as a function of the training size.}\label{fig:comparison_1}
\end{figure}

Given these results, the \texttt{Z15} and the \texttt{Z20} model were chosen to be compared with a classical linear support vector classifier (\texttt{SVC}). Figure~\ref{fig:comparison_1} shows a comparison of the performance of both models using $\mathrm{PPV}$, $\mathrm{TPR}$, $\mathrm{NPV}$, and $\mathrm{TNR}$ as performance metrics as a function of the training size. For most training sizes, the \texttt{Z20} quantum classifier presented a performance similar to the classical \texttt{SVC} classifier. It is worth noticing that the classical model performance collapsed for a training size of $90\%$ of the dataset.

\begin{figure}[htp]
\includegraphics[width=\textwidth]{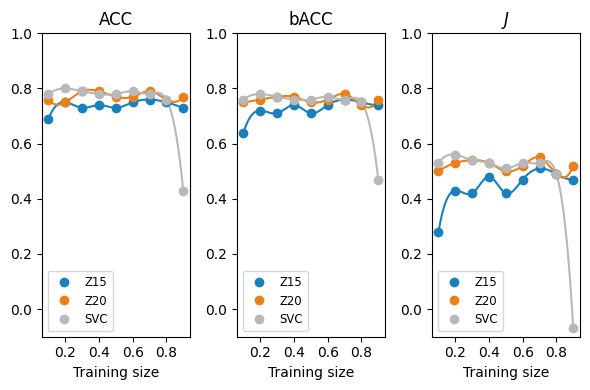}
\caption{Comparison of the performance of the \texttt{Z15}, \texttt{Z20} quantum models, and the \texttt{SVC} classical model as a function of the training size.}\label{fig:comparison_2}
\end{figure}

Figure~\ref{fig:comparison_2} presents accuracy $\mathrm{ACC}$, balanced accuracy $\mathrm{bACC}$, and Youden's index $J$ for the \texttt{Z15}, \texttt{Z20}, and the \texttt{SVC} classifiers. Performance of the \texttt{Z20} and the \texttt{SVC} models were similar, but the performance collapse of the classical classifier higher training size is clearly visible. In fact, the classical model exhibits $J\lessapprox 0$ for a training size of $90\%$, which is equivalent to a model without any predictive power. In contrast, the quantum models did not exhibit such a collapse in any of the experiments performed.

\section{Conclusion}\label{sec4}

In this paper, we presented an example of quantum machine learning applied to the well known case of the \texttt{titanic3} dataset classification. We built quantum models based on the Pauli expansion circuits \texttt{ZFeatureMap} and \texttt{ZZFeatureMap} as feature maps and the \texttt{RealAmplitudes} ansatz. In these models, the number of variational parameters of the ansatz ranged from $10$ to $50$.

The models were trained in a simulated hybrid quantum-classical setting, in which the classical optimizer COBYLA was used to optimize the variational parameters of the models that minimized the cross-entropy loss function. The models were assessed with several performance metrics that are widely used in classification tasks. The results showed that models based on the \texttt{ZFeatureMap} performed better than the models based on the \texttt{ZZFeatureMap}. The performance of the \texttt{ZFeatureMap}-based models seem to saturate above $15$ or $20$ variational parameters.

The \texttt{Z15} and \texttt{Z20} quantum classifiers were compared with a classical support vector classifier. This time, the classifiers were trained with a fraction of the dataset ranging from $10\%$ to $90\%$ of the data and tested on the remaining fraction of the data. The results indicate that the performance of the \texttt{Z20} quantum classifier is comparable with the classical classifier. Besides, the classical classifier collapsed when trained with $90\%$ of the data while the two quantum classifiers did not present any similar issue.

The results presented in this paper indicate that hybrid quantum-classical variational algorithms may be useful in classification tasks of classical data in the NISQ era.

\backmatter

\bmhead{Acknowledgements}
This study was financed, in part, by the S\~{a}o Paulo Research Foundation (FAPESP), Brasil, by the Coordena\c{c}\~{a}o de Aperfei\c{c}oamento de Pessoal de N\'{\i}vel Superior - Brasil (CAPES) - Finance Code 001, by the Conselho Nacional de Desenvolvimento Cient\'{\i}fico e Tecnol\'{o}gico (CNPq), and by the Funda\c{c}\~{a}o Educacional Inaciana Padre Sab\'{o}ia de Medeiros (FEI).

{}




\end{document}